\pdfoutput=1

\documentclass[11pt,twoside,a4paper,cmspaper,final,collab]{cms-tdr}
\def\svnVersion{257861}\def\cmsCernNo{2013-037}\def\cmsCernDate{\today}\def\cmsMessage{Published in the Journal of High Energy Physics as \href{http://dx.doi.org/10.1007/JHEP04(2014)191}{\doi{10.1007/JHEP04(2014)191.}}}

\begin{document}\cmsNoteHeader{TOP-12-010}

\hyphenation{had-ron-i-za-tion}
\hyphenation{cal-or-i-me-ter}
\hyphenation{de-vices}

\cmsNoteHeader{TOP-12-010} 
\title{Measurements of the \ttbar charge asymmetry using the dilepton decay channel in pp collisions at $\sqrt{s}=7$\TeV}

\date{\today}
\definecolor{light-gray}{gray}{0.30}

\newcommand{\ttfake}{\ttbar\ (non-dileptonic)}
\newcommand{\wjets}{$\PW+\text{jets}$}

\providecommand{\fig}{Fig.}
\providecommand{\figs}{Figs.}
\providecommand{\tab}{Table}
\providecommand{\tabs}{Tables}
\providecommand{\secn}{Section}
\providecommand{\secns}{Sections}
\providecommand{\reference}{Ref.}
\providecommand{\references}{Refs.}

\abstract{The \ttbar  charge asymmetry in proton-proton collisions at $\sqrt{s}=7$\TeV is measured using the dilepton decay channel (ee, $\Pe\mu$, or $\Pgm\Pgm$). The data correspond to a total integrated luminosity of 5.0\fbinv, collected by the CMS experiment at the LHC. The \ttbar  and lepton charge asymmetries, defined as the differences in absolute values of the rapidities between the reconstructed top quarks and antiquarks and of the pseudorapidities between the positive and negative leptons, respectively, are measured to be $A_\mathrm{C} = -0.010 \pm 0.017\stat \pm 0.008\syst$ and $A^\text{lep}_\mathrm{C} = 0.009 \pm 0.010\stat \pm 0.006\syst$. The lepton charge asymmetry is also measured as a function of the invariant mass, rapidity, and transverse momentum of the \ttbar  system. All measurements are consistent with the expectations of the standard model.}

\hypersetup{%
pdfauthor={CMS Collaboration},%
pdftitle={Measurements of the t t-bar charge asymmetry using the dilepton decay channel in pp collisions at sqrt(s) = 7 TeV},%
pdfsubject={CMS},%
pdfkeywords={CMS, physics, top, charge, asymmetry}}

\maketitle 

\section{Introduction}
\label{sec:intro}

Among the standard model (SM) fermions, the top quark is distinguished by its large mass.
In several theories of physics beyond the SM, new phenomena are predicted through interactions involving top quarks.
Measuring the properties of top quarks is therefore important not only for checking the validity of the SM, but also as a key probe of possible new physics.
Recent measurements of the \ttbar\ forward-backward production asymmetry ($A_{\mathrm{fb}}$) from the D0~\cite{ref:D0afb}
and CDF~\cite{ref:CDFafb2013} experiments at the Tevatron indicate possible disagreement with SM expectations,
particularly at large \ttbar\ invariant mass.

Unlike the Tevatron proton-antiproton collider,
the Large Hadron Collider (LHC) is a proton-proton collider,
which lacks a natural definition for the charge asymmetry given
the symmetric nature of the incoming protons.
However, the parton distributions inside the protons are not
symmetric for quarks (mainly valence quarks) and antiquarks (all sea quarks),
meaning
quarks ($\cPq$) usually carry more momentum than antiquarks ($\cPaq$).
For a positive (negative) charge asymmetry in $\cPq\cPaq \to \ttbar$ events, the top quark (top antiquark) is more likely to be produced in the direction of the incoming quark in the \ttbar\ rest frame,
resulting in a broader (narrower) rapidity distribution of top quarks than of top antiquarks in the laboratory frame.
The difference in the absolute values of the rapidities ($y$) of the top quarks and antiquarks, $\Delta \abs{y_\cPqt} = \abs{y_\cPqt} - \abs{y_{\cPaqt}}$,
is therefore a suitable observable to measure the \ttbar\ charge asymmetry $A_\mathrm{C}$, defined as

\begin{equation*}
A_\mathrm{C} = \frac{N(\Delta \abs{y_\cPqt} > 0)-N(\Delta \abs{y_\cPqt} < 0)}{N(\Delta \abs{y_\cPqt} > 0)+N(\Delta \abs{y_\cPqt} < 0)}.
\end{equation*}

A similar observable~\cite{ref:Krohn} involving the difference in the absolute values of the pseudorapidities ($\eta$, to be defined in the next section) of the positive and negative leptons in dileptonic \ttbar\ events, $\Delta \abs{\eta_\ell} = |\eta_{\ell^+}| - |\eta_{\ell^-}|$,
is used to define the lepton charge asymmetry:

\begin{equation*}
A^\text{lep}_\mathrm{C} = \frac{N (\Delta \abs{\eta_\ell} > 0)-N (\Delta \abs{\eta_\ell} < 0)}{N (\Delta \abs{\eta_\ell} > 0)+N (\Delta \abs{\eta_\ell} < 0)}.
\end{equation*}

In the SM, a small positive charge asymmetry arises from corrections to the tree-level $\cPq\cPaq \to \ttbar$ process, as explained in detail in \reference~\cite{ref:BernChargeAsym}.
There are models of new physics that predict larger values of $A_{\mathrm{fb}}$ than
expected in the SM from the interference of SM \ttbar\ production with contributions from processes such as
$s$-channel axigluon or $t$-channel \PWpr or \PZpr exchange~\cite{ref:Krohn}.
Such theories predict values of $A_\mathrm{C}$ and $A^\text{lep}_\mathrm{C}$ over a large range~\cite{ref:Krohn}, and
accurate measurements of these quantities can therefore provide important constraints.

This paper presents the first measurements of $A_\mathrm{C}$ and $A^\text{lep}_\mathrm{C}$ in the dilepton final state,
using data from pp collisions at $\sqrt{s}=7$\TeV,
corresponding to an integrated luminosity of 5.0\fbinv recorded by the Compact Muon Solenoid (CMS) experiment at the LHC.
Previously, using a single-lepton \ttbar
event sample, CMS determined $A_\mathrm{C} = 0.004 \pm 0.010\stat \pm 0.011\syst$~\cite{semileptonicCMS},
while the ATLAS Collaboration measured $A_\mathrm{C} = 0.006 \pm 0.010\,(\text{total})$~\cite{ref:ATLASchargeasym,Aad:2013cea}, both
consistent with the SM
prediction of $A_\mathrm{C} = 0.0123 \pm 0.0005$~\cite{ref:BernChargeAsym}.

The analysis described in this paper uses a complementary data
sample to that used in \reference~\cite{semileptonicCMS}. The \ttbar\ dilepton decay
channel has a smaller background than the single-lepton channel
and different systematic uncertainties.  Furthermore, the
dilepton channel allows us to measure the lepton charge asymmetry
$A^\text{lep}_\mathrm{C}$ for the first time.
The SM prediction for $A^\text{lep}_\mathrm{C}$ is $0.0070 \pm 0.0003$~\cite{ref:BernChargeAsym}.
We also measure $A^\text{lep}_\mathrm{C}$
differentially as a function of
three variables describing the \ttbar\ system in the laboratory frame: its invariant mass ($M_{\ttbar}$), rapidity ($\abs{y_{\ttbar}}$), and transverse momentum ($\pt^{\ttbar}$).
Since the reconstructed asymmetries are distorted by detector effects, we apply an unfolding technique to determine the parton-level distributions, which can be directly compared with
theoretical predictions.

\section{CMS detector}

The central feature of the CMS apparatus is a superconducting solenoid,
13~m in length and 6\unit{m} in diameter, which provides an axial magnetic
field of 3.8\unit{T}.  The bore of the solenoid is equipped with a variety of
particle detection systems. Charged-particle trajectories are
measured with a silicon pixel and strip tracker,
covering $0 \leq \phi < 2\pi$ in azimuth and
the pseudorapidity region
$\abs{\eta}<2.5$, where $\eta =-\ln[\tan{\theta/2}]$ with $\theta$ the
polar angle of the trajectory of the particle with respect to the anticlockwise-beam direction.
A crystal electromagnetic calorimeter and a brass/scintillator hadron calorimeter surround
the silicon tracking volume and provide high-resolution measurements of energy used to
reconstruct electrons, photons, and jets. Muons are measured in gas-ionisation detectors embedded in the
steel flux return yoke of the solenoid. The detector is nearly hermetic, thereby providing reliable
measurements of momentum imbalance in the plane transverse to the beams.
A trigger system selects the most interesting collisions for analysis.
A more detailed description of the CMS detector is given in \reference~\cite{JINST}.

\section{Event samples, reconstruction, and selection}
\label{sec:eventsel}

Events are selected using triggers that require the presence of at least two leptons (electrons or muons) with transverse momentum (\pt) requirements of $\ge17\GeV$ for the highest-\pt lepton and $\ge8\GeV$ for the second-highest-\pt lepton.
Electron candidates~\cite{EGMPAS} are reconstructed by associating tracks from the silicon tracker with energy clusters in the electromagnetic
calorimeter. Muon candidates~\cite{MUOART} are reconstructed by combining information from the muon detector with tracks reconstructed in the silicon tracker.
Additional lepton identification criteria are applied to both lepton flavours in order to reject hadronic jets misreconstructed as leptons~\cite{EGMPAS,MUOART}.
Both electrons and muons are required to be isolated from other activity in the event.
This is achieved by imposing a maximum value of 0.15 on the relative isolation of the leptons.
This is defined as the scalar sum of all additional silicon track \pt and calorimeter transverse energy (energy deposits projected onto the plane transverse to the beam)
within a cone of ${ \Delta R}\equiv\sqrt{\smash[b]{{ (\Delta\eta)^2+(\Delta\phi)^2}}}=0.3$ around the lepton candidate direction,
divided by the lepton candidate \pt~\cite{ref:1105.5661}.
Here, $\Delta\eta$ and $\Delta\phi$ are the differences in pseudorapidity and azimuthal angle between the lepton candidate and the additional track or calorimeter energy deposit.

Selections are applied to reject events other than from \ttbar\ production in the dilepton final state.
Events are required to contain two isolated leptons of opposite electric charge ($\Pep\Pem$, $\Pe^{\pm}\mu^{\mp}$, or $\Pgmp\Pgmm$).
The electrons and muons are required to have $\pt >  20\GeV$ and $\abs{\eta} < 2.5$ and $2.4$, respectively.
The two reconstructed lepton trajectories must be consistent with originating from a common interaction vertex.
Events with an $\Pep\Pem$  or $\Pgmp\Pgmm$ pair having an  invariant mass in the \cPZ-boson mass ``window'' (between 76 and  106\GeV) or below
20\GeV are removed to suppress Drell--Yan (\ensuremath{\cPZ/\Pgg^\star}+jets) and heavy-quarkonium resonance production.
The jets and the transverse momentum imbalance in each event are reconstructed using a particle-flow technique~\cite{CMS-PAS-PFT-10-002}.
The  anti-\kt clustering
algorithm~\cite{antikt} with a distance parameter of 0.5 is used for jet
clustering.
Corrections are applied to the energies of the reconstructed jets, based on the results of
simulations and studies using exclusive dijet and $\gamma$+jets data~\cite{Chatrchyan:2011ds}.
At least two jets with  $\pt > 30\GeV$ and  $\abs{\eta} < 2.5$,
separated  by $\Delta  R  >  0.4$ from the leptons  that pass the  analysis
selection, are required in each event.
At least one of these jets must be
consistent with the decay of a heavy-flavour hadron (a ``\cPqb~jet''),
identified
by the Combined Secondary Vertex \cPqb-tagging algorithm~\cite{ref:btag}.
This algorithm is based on the reconstruction of a secondary decay vertex,
and an operating point is chosen that
gives a \cPqb-tagging efficiency of about 70\% (depending on jet \pt\ and $\eta$)
with misidentification probabilities of approximately 1.5\% and 20\% for jets originating from light partons (\cPqu, \cPqd, and \cPqs\ quarks, and gluons) and \cPqc~quarks, respectively.
The missing transverse energy in an event, \MET, is defined as the magnitude of
the transverse momentum imbalance, which is the negative of the vector sum of the \pt\ of all
reconstructed particles.
The  \MET\ value is required to exceed 40\GeV in events with same-flavour leptons in order to further suppress the Drell--Yan background.
There is no \MET\ requirement for $\Pe^{\pm}\Pgm^{\mp}$ events.

Simulated \ttbar\ events are generated using the \MCATNLO3.41~\cite{mc@nlo} Monte Carlo generator,
with a top-quark mass of $m_\cPqt=172.5$\GeV,
and the parton showering and fragmentation performed using \HERWIG6.520~\cite{herwig6}.
Simulations with different values of $m_\cPqt$ and factorisation and renormalisation scales are used to evaluate the associated systematic uncertainties.
Background samples of \wjets, Drell--Yan, diboson ($\PW\PW$, $\PW\cPZ$, and $\cPZ\cPZ$), and single-top-quark events are generated with \MADGRAPH~\cite{Alwall:2011uj} or \POWHEG~\cite{Nason:2004rx,Frixione:2007vw,Alioli:2010xd},
and the parton showering and fragmentation is done using \PYTHIA6.4.22~\cite{Pythia}.
Next-to-leading-order (NLO) or next-to-next-to-leading-order cross sections are used to normalise the background samples~\cite{Melnikov:2006kv,Alioli:2008gx,Campbell:2011bn,Kidonakis:2011wy,Kidonakis:2010tc,Alioli:2009je,Kidonakis:2010ux,Re:2010bp}.

For both signal and background events, additional \Pp\Pp\ interactions in the same or nearby bunch crossings (``pileup'') are simulated with \PYTHIA\ and superimposed on the hard collisions,
using a pileup multiplicity distribution that reflects the luminosity profile of the analysed data.
The CMS detector response is simulated using a \GEANTfour-based model~\cite{Geant}.
The simulated events are reconstructed and analysed with the same software used to process the data.

The trigger efficiency for dilepton events that satisfy the selection criteria is determined using a tag-and-probe method~\cite{wzPAS2010}.
The efficiencies for the $\Pe\Pe$, $\Pe\Pgm$, and $\Pgm\Pgm$ channels are approximately 100\%, 95\%, and 90\%, respectively, each with an uncertainty of about 2\%~\cite{ref:tprime}.
These efficiencies are used to weight the simulated events to account for the trigger requirement.
The lepton selection efficiencies (reconstruction, identification, and isolation) are consistent between data and simulation~\cite{wzPAS2010, PAPER-TOP-11-005}.
To account for the differences between b-tagging efficiencies measured in data and simulation~\cite{ref:btag}, data-to-simulation scale factors are applied for each jet in simulated events.
Previous CMS studies~\cite{toppT} have shown that the \pt\ distribution of the top quark in data is softer than in the NLO simulation. Reweighting the top-quark \pt\ spectrum in the simulation to match the data
improves the modelling of the lepton and jet \pt\ distributions, and is applied to the \MCATNLO\ \ttbar\ sample.

\section{Background estimation}
\label{Sec:BkgEst}

The backgrounds from events with a jet misidentified as a lepton and from Drell--Yan production are estimated using both data- and simulation-based techniques.
The results agree within their uncertainties.
The simulation is chosen as the method to predict the yields and distributions of the backgrounds, with systematic uncertainties
based on a comparison with the data-based estimates.
Contributions to the background from single-top-quark and diboson events are estimated from simulation alone.
Recent measurements from the CMS Collaboration~\cite{ref:singletop,ref:WWWZ} indicate agreement between the predicted and measured cross sections for these processes.

The background with at least one misidentified lepton (non-dileptonic \ttbar, \wjets, and multijet events) is estimated
from data using a \pt- and $\eta$-dependent parameterisation of the probability for a jet to be misidentified as a lepton, determined
using events collected with jet triggers of different energy thresholds~\cite{ref:samesign}. For both the electron and muon candidates described in \secn~\ref{sec:eventsel}, an associated ``loose" lepton candidate is defined based on relaxed isolation requirements~\cite{ref:samesign}. The lepton misidentification probabilities
are then applied as weights to events containing one lepton candidate passing the signal selection and one or more loose lepton candidates.

The Drell--Yan background outside the Z-boson mass window is estimated using the ratio of the numbers of simulated events inside and outside the window to scale
the observed event yield inside the window~\cite{ref:1105.5661}.
Contributions to this region from other processes, in which the two leptons do not arise from Z-boson decay, are
estimated from the number of $\Pe\Pgm$ events in data and subtracted prior to performing the rescaling.

\section{Event yields and measurements at the reconstruction level}
\label{sec:yields}

The expected background and observed event yields per lepton flavour combination in the final sample are listed in \tab~\ref{tab:yields1}.
The total predicted yield in the $\Pe\Pgm$ channel is significantly larger than for the same-flavour channels, for which the additional requirements on the \MET\ and invariant-mass of the lepton pair described in \secn~\ref{sec:eventsel} are applied to suppress Drell--Yan background.
After subtraction of the predicted background yields, the remaining yield in data is assumed to be signal from dileptonic \ttbar\ decays,
including $\tau$ leptons that decay leptonically. All other \ttbar\ decay modes are treated as background and are included in the non-dileptonic \ttbar\ category.
The largest background comes from single-top-quark production.
The systematic uncertainties in the simulated yields are discussed in \secn~\ref{sec:systematics}.

\begin{table*}[htb]
\centering
\topcaption{\label{tab:yields1}
The predicted background and observed event yields after applying the event selection criteria and normalisation described in the text.
Uncertainties are statistical only.}
\setlength{\extrarowheight}{1.5pt}
\begin{tabular}{l |  c  c  c  c}
\hline
\multicolumn{1}{c|}{Sample}  & $\Pe\Pe$ & $\Pgm\Pgm$ & $\Pe\Pgm$ & All \\
\hline
    \ttfake\ &  38.3 $\pm$   1.6 &   4.02 $\pm$   0.45 &  91.7 $\pm$   2.4 & 134.0 $\pm$   2.9 \\
    \wjets\   &   ${<}2.0$    &   4.7 $\pm$   3.3 &  11.1 $\pm$   5.1 &  15.8 $\pm$   6.1 \\
    Drell--Yan &  30.2 $\pm$   4.4 &  29.6 $\pm$   4.1 &  35.0 $\pm$   4.5 &  94.8 $\pm$   7.5 \\
    Diboson   &   8.27 $\pm$   0.44 &  10.20 $\pm$   0.47 &  27.90 $\pm$   0.81 &  46.4 $\pm$   1.0 \\
    Single top-quark  &  72.5 $\pm$   2.1 &  86.8 $\pm$   2.2 & 289.4 $\pm$   4.2 & 448.7 $\pm$   5.2 \\
    \hline
 Total (background) & 149.3 $\pm$  5.5 & 135.3 $\pm$  5.8 & 455.1 $\pm$  8.4 & 740 $\pm$  11 \\  \hline
     Data   & 1631 & 1964 & 6229 & 9824 \\  \hline
\end{tabular}
\end{table*}

The measurement of the \ttbar\ charge asymmetry using $\Delta \abs{y_\cPqt}$ requires the reconstruction of the entire \ttbar\ event.
Each signal event contains two neutrinos, and there is also an ambiguity in combining the b~jets with the leptons, resulting in up to 8 possible solutions for the \ttbar\ system.
The Analytical Matrix Weighting Technique (AMWT)~\cite{ref:1105.5661} is used to find the most probable solution
for a top-quark mass of $172.5$\GeV. In events with only one b~tag, the second b~jet is assumed to be the untagged jet with the largest \pt.
Solutions are assigned a weight based on the probability
of observing the given configuration~\cite{ref:1105.5661}, and the \ttbar\ kinematic quantities are taken from the solution with the largest weight. To reduce the fraction of events with no analytic solution, caused largely by the presence
of mismeasured jets,
the \MET\ and the energies and directions of the jets are allowed to vary within their uncertainties
via a Monte Carlo integration over parameterised jet and \MET\ resolution functions~\cite{Chatrchyan:2011ds}.
Despite this step, ${\approx}14\%$ of the events still provide no solutions,
both for data and simulation.
In the measurement of $\Delta \abs{y_\cPqt}$, $M_{\ttbar}$, $\abs{y_{\ttbar}}$, and $\pt^{\ttbar}$,
 these events are not used, which
is accounted for as an additional event selection requirement.

\begin{figure}[!htpb]
\centering
\includegraphics[width=0.48\linewidth]{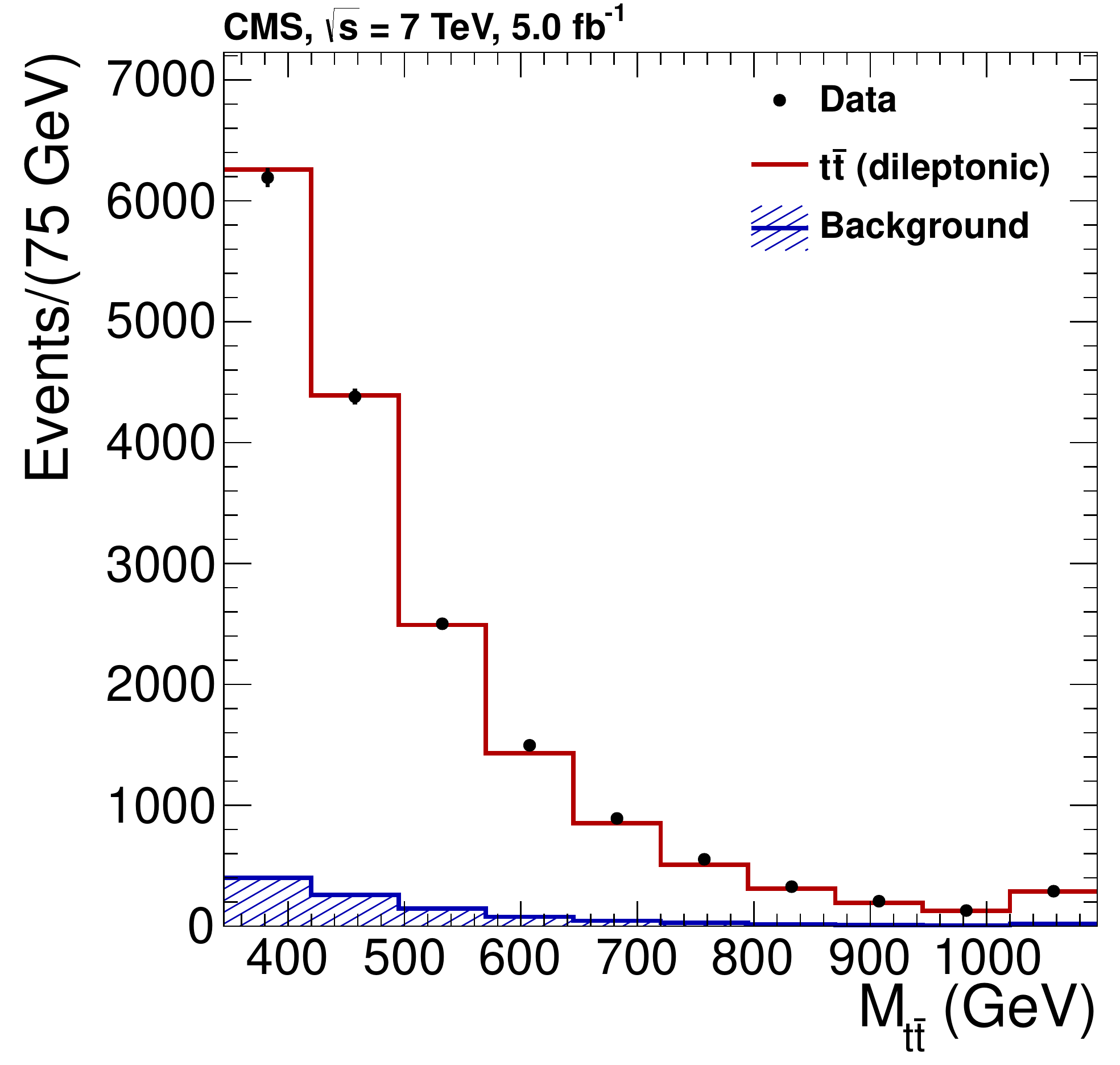}

\includegraphics[width=0.48\linewidth]{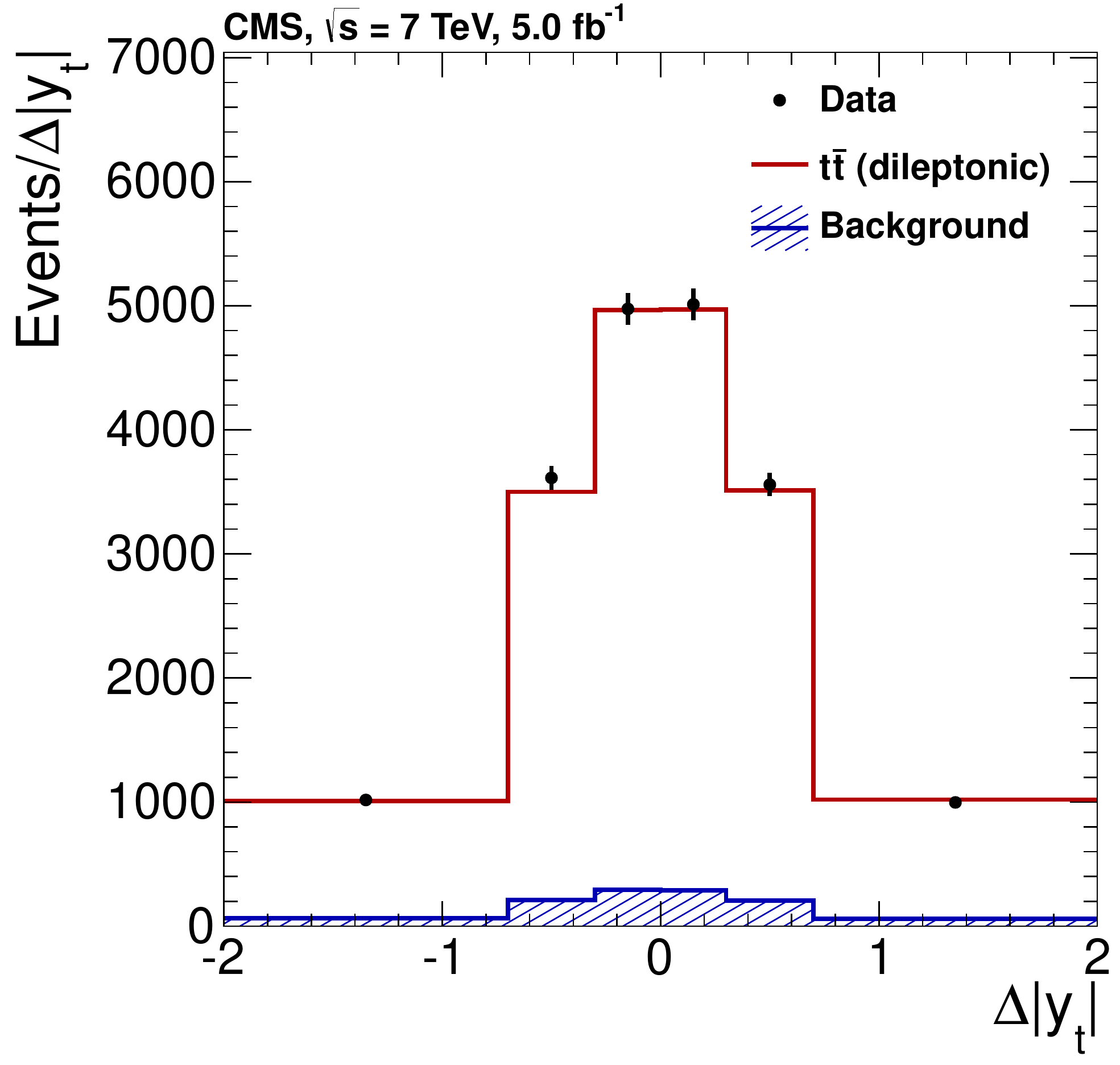}
\includegraphics[width=0.48\linewidth]{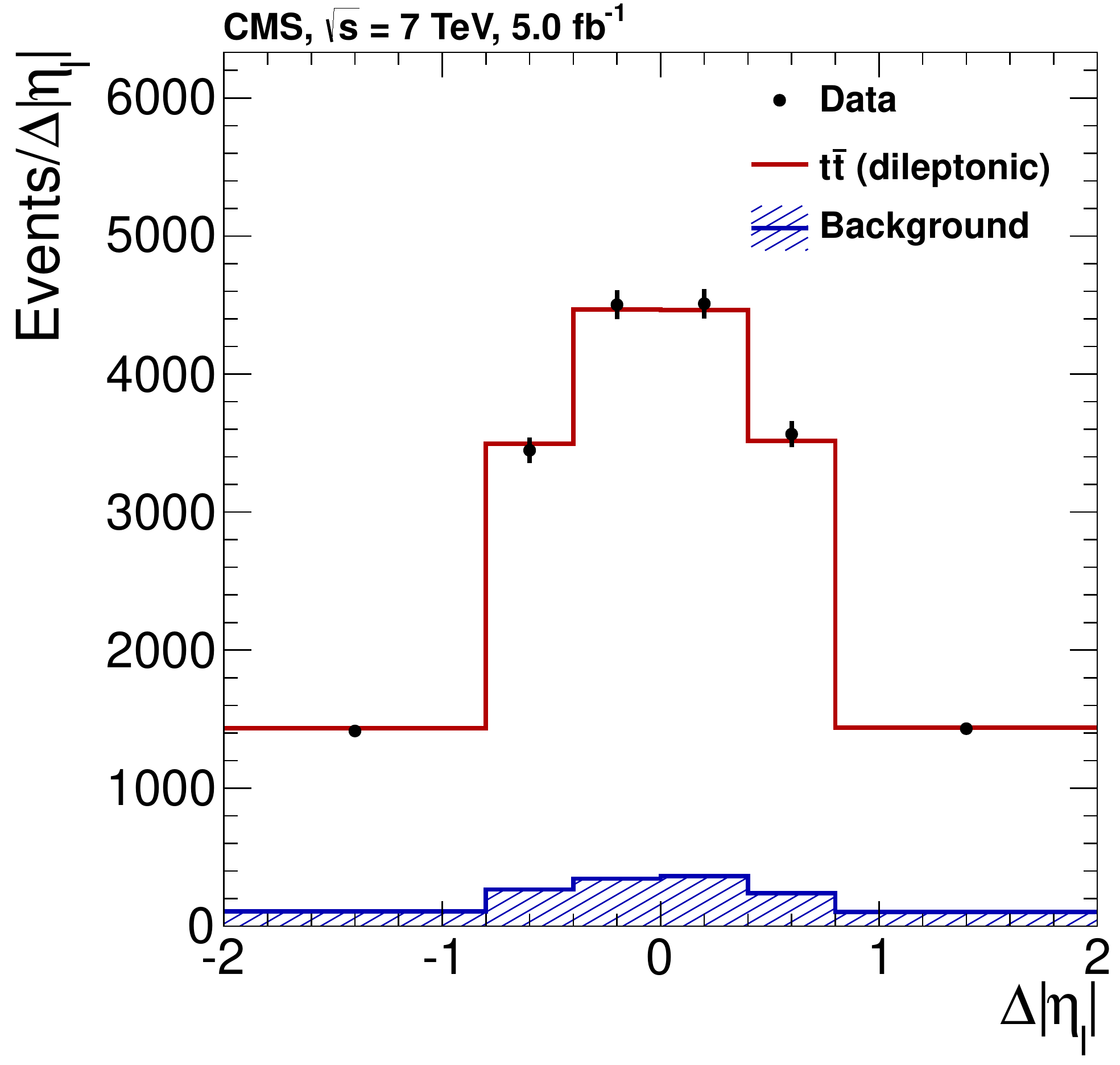}
\caption{\label{fig:preselcomp}\protect
The reconstructed $M_{\ttbar}$ (top), $\Delta \abs{y_\cPqt}$ (bottom left), and $\Delta \abs{\eta_\ell}$ (bottom right) distributions from data (points) and simulation (histogram).
The simulated events are divided into signal (open histogram) and background (dashed histogram) contributions, where the background contribution includes all event categories stipulated in \tab~\ref{tab:yields1}.
The signal yield is normalised to the background-subtracted data.
The first and last bins include underflow and overflow events, respectively.
The error bars on the data points represent the statistical uncertainties only.
}
\end{figure}

A comparison between data and simulation for the $M_{\ttbar}$, $\Delta \abs{y_\cPqt}$, and $\Delta \abs{\eta_\ell}$ distributions is shown in \fig~\ref{fig:preselcomp},
where the signal yield from the simulation has been normalised to the number of background-subtracted signal events in data.
The distributions from data and simulation agree in all cases.
The uncorrected value of $A_\mathrm{C}$ at the reconstruction level is  $-0.005 \pm 0.011$  in data and $0.003 \pm 0.003$ in simulation, where the uncertainties are statistical only.
For $A^\text{lep}_\mathrm{C}$, the uncorrected values are $0.007 \pm 0.010$ and $0.002 \pm 0.003$ in data and simulation, respectively.

\section{Unfolding the distributions}
\label{sec:unfolding}
The observed  $\Delta \abs{y_\cPqt}$ and $\Delta \abs{\eta_\ell}$  distributions are distorted relative to the true underlying distributions
by the acceptance of the detector, the efficiency of the trigger and event selection, and the finite
resolution of the kinematic quantities. To correct the data for these effects, we apply an unfolding procedure that
 yields the corrected $\Delta \abs{y_\cPqt}$ and $\Delta \abs{\eta_\ell}$ distributions at the parton level.
These distributions represent the differential cross sections in $\Delta \abs{y_\cPqt}$ and $\Delta \abs{\eta_\ell}$, and are normalised to unit area.

The choice of binning for each distribution is motivated by the desire to minimise bin-to-bin statistical
fluctuations.
The bin sizes are chosen so that there are similar numbers of events in each bin, and are summarised in \tab~\ref{tab:Binning}.

\begin{table}[!htpb]
\centering
\topcaption{\label{tab:Binning} Binning used in the distributions of $\Delta \abs{y_\cPqt}$ and $\Delta \abs{\eta_\ell}$.}
\begin{tabular}{c  c  c  c  c  c  c }
\hline
$\Delta \abs{y_\cPqt}$ & [$-\infty$, $-0.7$]  &  [$-0.7$, $-0.3$]  &  [$-0.3$, $0.0$]  &  [0.0, 0.3]  &  [0.3, 0.7]  &  [0.7, $\infty$] \\
$\Delta \abs{\eta_\ell}$ &   [$-\infty$, $-0.8$]  &  [$-0.8$, $-0.4$]  &  [$-0.4$, $0.0$]  &  [0.0, 0.4]  &  [0.4, 0.8]  &  [0.8, $\infty$] \\ \hline
\end{tabular}
\end{table}

\begin{figure}[!htpb]
\centering
\includegraphics[width=0.48\linewidth]{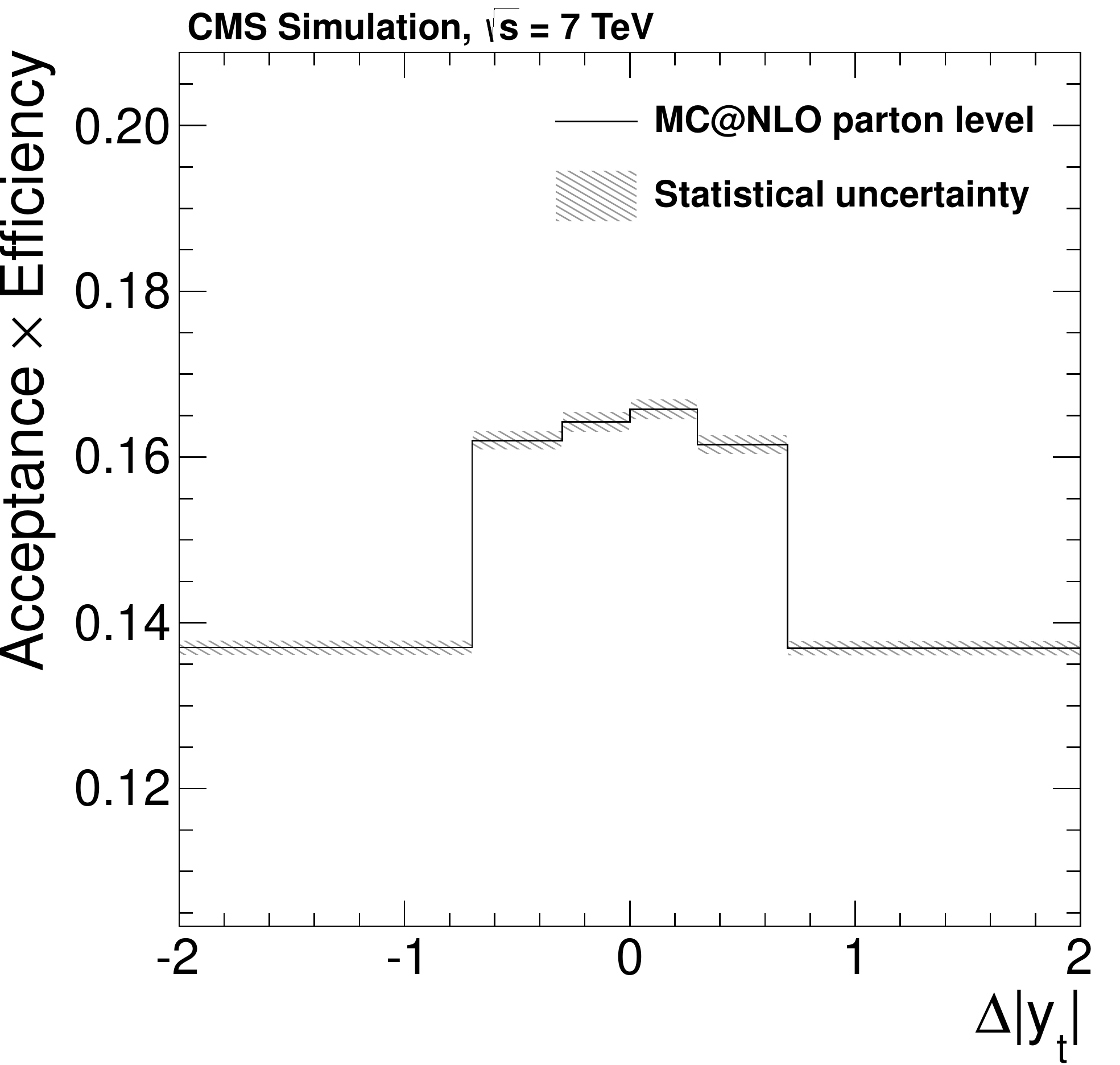}
\includegraphics[width=0.48\linewidth]{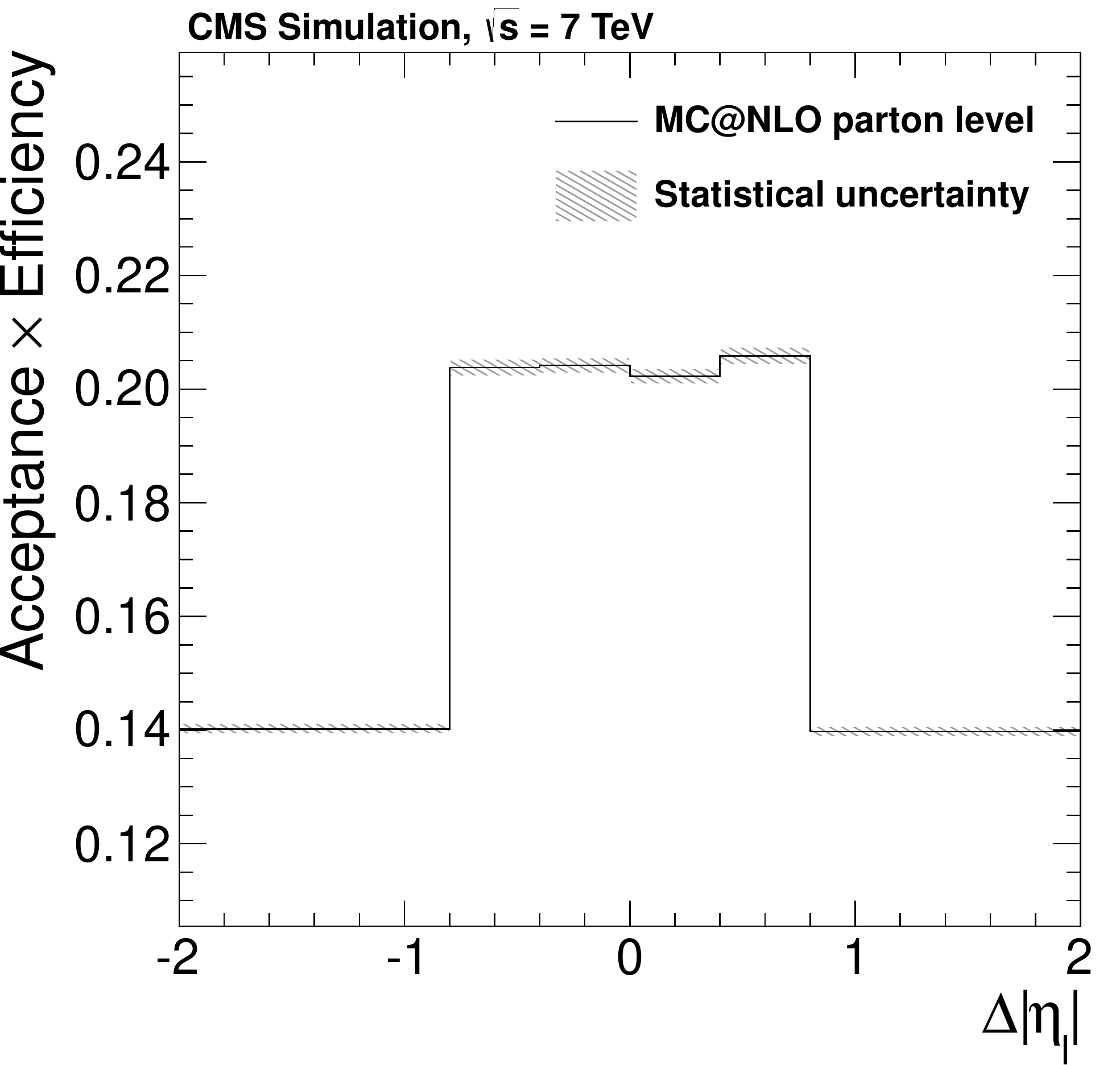}
\caption{\label{fig:AcceptMatrix}\protect
Diagonal elements of the matrix $A$ describing the acceptance times efficiency of signal events as a function of $\Delta \abs{y_\cPqt}$ (left) and $\Delta \abs{\eta_\ell}$ (right) from simulated \MCATNLO\ \ttbar\ events.
The statistical uncertainties are represented by the hatched band, and
the first and last bins include underflow and overflow events, respectively.
}
\end{figure}

\begin{figure}[!htpb]
\centering
\includegraphics[width=0.48\linewidth]{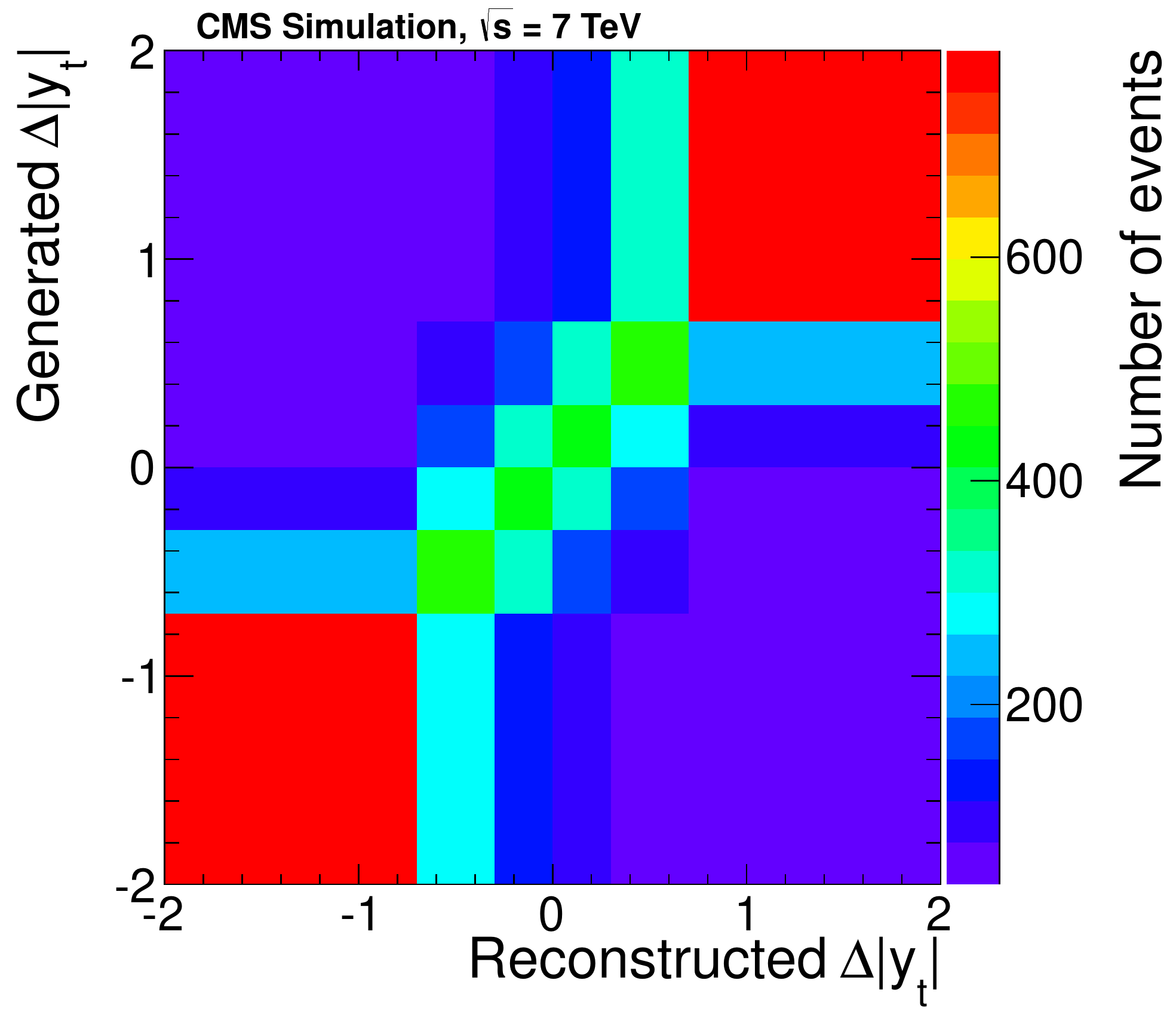}
\includegraphics[width=0.48\linewidth]{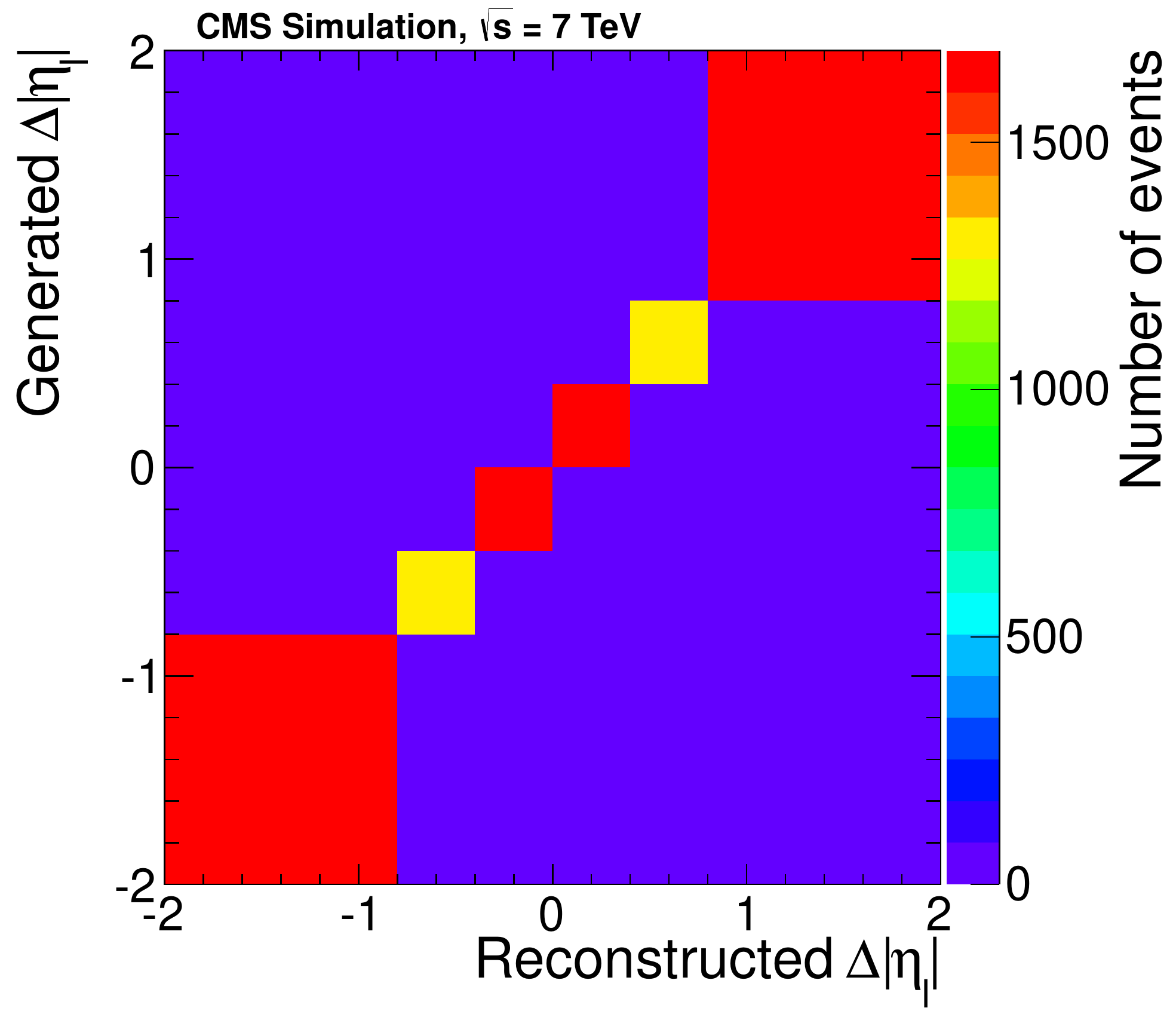}
\caption{\label{fig:SmearMatrix}\protect
Binned distributions of generated versus reconstructed values of $\Delta \abs{y_\cPqt}$ (left) and $\Delta \abs{\eta_\ell}$ (right) from simulated \MCATNLO\ \ttbar\ events, used to derive the smearing matrices ($S$).
}
\end{figure}

The background-subtracted distribution $\vec{b}$ for either $\Delta \abs{y_\cPqt}$ or $\Delta \abs{\eta_\ell}$ is related to the underlying
parton-level distribution $\vec{x}$ through the equation $\vec{b}=SA\vec{x}$,
where $A$ is a diagonal matrix describing the fraction (acceptance times efficiency) of all produced signal events that are expected to be selected in each of the measured bins, and
$S$ is a non-diagonal ``smearing'' matrix describing the migration of events between bins caused by the detector resolution
and reconstruction techniques. The $A$ and $S$ matrices are modelled using simulated \MCATNLO\ \ttbar\ events,
and the results are displayed in \figs~\ref{fig:AcceptMatrix} and~\ref{fig:SmearMatrix}.
The smearing in $\Delta \abs{y_\cPqt}$ can be large in some events because of the uncertainties in the reconstruction of the \ttbar\ kinematic quantities.
However, the largest numbers of events in the left plot of \fig~\ref{fig:SmearMatrix} lie close to the diagonal, meaning there is little
migration between bins that are far apart.
The corresponding smearing matrix for $\Delta \abs{\eta_\ell}$, shown in the right plot of \fig~\ref{fig:SmearMatrix}, is close to diagonal because of the excellent angular resolution of the lepton measurements.

To determine the parton-level distributions for $\Delta \abs{y_\cPqt}$ and $\Delta \abs{\eta_\ell}$,
we employ a regularised unfolding algorithm based on singular-value decomposition (SVD)~\cite{Hocker:1995kb}.
 The effects of large statistical fluctuations in the algorithm are
greatly reduced by introducing a regularisation term in the unfolding procedure.
The full covariance matrix is used in the evaluation of the statistical uncertainty in the measured asymmetry.

To verify that the unfolding procedure correctly unfolds distributions for different values of the asymmetry,
we reweight simulated \ttbar\ events according to a linear function of $\Delta \abs{y_\cPqt}$ (or $\Delta \abs{\eta_\ell}$), defined by a weight $w=1+K \Delta \abs{y_\cPqt}$ (or $\Delta \abs{\eta_\ell}$).
The parameter $K$ is varied between $-0.3$ and 0.3 in steps of 0.1, introducing asymmetries between approximately $-0.2$ and $0.2$ (far beyond the SM expectations).
For each value of $K$, we generate a set of pseudoexperiments
in which the number of events in each bin of the measured distribution is varied according to Poisson
statistics. The distributions are then unfolded, and the average value of the measured asymmetry is
compared to the input value.
We observe a linear relationship, thus validating the unfolding procedure.
The constant of proportionality between the true and measured asymmetries deviates slightly from unity, leading to changes of up to 1$\%$ in the measured asymmetry.
The effect of this bias is included in the systematic uncertainty from the unfolding.
We also fit the distribution of the pulls ([measured-expected]/uncertainty) in the set of pseudoexperiments to a Gaussian function and verify that its standard deviation is consistent with unity.

\section{Systematic uncertainties}	
\label{sec:systematics}

Various systematic uncertainties have been evaluated, concerning mainly the detector performance and the modelling of the signal and background processes.
Each systematic uncertainty is estimated using the difference between the results from the systematic variation and the central value.

The uncertainty from the jet-energy-scale (JES) corrections affects the AMWT \ttbar\ solutions, as well as the event selection.
It is estimated by varying the JES of jets within their uncertainties~\cite{Chatrchyan:2011ds}, and propagating this to the \MET.
The uncertainty in the lepton energy scale, which affects mainly the lepton \pt\ distributions, is estimated by varying the energy scale of electrons by ${\pm}0.5\%$ (the uncertainty in muon energies is negligible in comparison), as estimated from comparisons between measured and simulated \cPZ-boson events~\cite{1748-0221-8-09-P09009}.

The uncertainty in the background subtraction is obtained by
varying the normalisation of each background component, by ${\pm}50\%$ for single-top-quark and diboson production, and by ${\pm}100\%$ for Drell--Yan production and misidentified leptons, based on the estimates discussed in \secn~\ref{Sec:BkgEst}.

The \ttbar\ modelling and simulation uncertainties are evaluated by rederiving the $A$ and $S$ matrices
using simulated events with the following variations:
the jet energy resolution is increased by 5--10\%, depending on the $\eta$ of the jet~\cite{Chatrchyan:2011ds};
the simulated pileup multiplicity distribution is changed within its uncertainty;
the scale factors between data and simulation for the \cPqb-tagging efficiency~\cite{ref:btag}, trigger efficiency, and lepton selection efficiency are shifted up and down by their uncertainties;
the factorisation and renormalisation scales are together varied up and down by a factor of 2;
the top-quark mass is varied by $\pm$1\GeV, based on the uncertainty in the combined Tevatron $m_\cPqt$ measurement~\cite{Aaltonen:2012ra};
and the parton distribution functions are varied using the {\sc{pdf4lhc}} formula~\cite{pdf4lhcInterim}.
In the simulated \ttbar\ events, the $\tau$-leptons are unpolarised. This affects the angular distributions of the electrons and muons coming from $\tau$-lepton decays. The corresponding systematic effect is estimated by reweighting the $\tau$-lepton decay distributions to reproduce the SM expectations.
Since the origin of the discrepancy of the top-quark \pt\ distributions between data and simulation~\cite{toppT} is not fully understood,
a 100\% systematic uncertainty is applied to the top-quark \pt\ reweighting procedure discussed in \secn~\ref{sec:eventsel}.

Finally, the results of the unfolding linearity tests discussed in \secn~\ref{sec:unfolding} are used to estimate the systematic uncertainty in the unfolding procedure.
The systematic uncertainties in the unfolded $A_\mathrm{C}$ and $A^\text{lep}_\mathrm{C}$ measurements are summarised in \tab~\ref{tab:asyms_sys}.
The individual terms are added in quadrature to estimate the total systematic uncertainties.
The dominant uncertainties are from the unfolding procedure for $A_\mathrm{C}$, and the factorisation and renormalisation scale uncertainties for $A^\text{lep}_\mathrm{C}$.

\begin{table}[!h]						
\centering	
\topcaption{Systematic uncertainties in the unfolded values of $A_\mathrm{C}$ and $A^\text{lep}_\mathrm{C}$ from the sources listed.}
\label{tab:asyms_sys}							
\begin{tabular}{l|rr}						
\hline & & \\ [-2.2ex]
Variable		&	\multicolumn{1}{c}{$A_\mathrm{C}$}	&	\multicolumn{1}{c}{$A^\text{lep}_\mathrm{C}$}	\\ [0.3ex]
\hline														
\multicolumn{3}{c}{Experimental uncertainties}	\\						
\hline							
Jet energy scale		&$	0.003	$&$	0.001	$	\\
Lepton energy scale		&$	{<}0.001	$&$	{<}0.001	$	\\
Background		&$	0.001	$&$	0.001	$	\\
Jet energy resolution		&$	{<}0.001	$&$	{<}0.001	$	\\
Pileup		&$	{<}0.001	$&$	0.001	$	\\
Scale factor for b tagging	&$	{<}0.001	$&$	{<}0.001	$	\\
Lepton selection		&$	{<}0.001	$&$	{<}0.001	$	\\
\hline														
\multicolumn{3}{c}{\ttbar\ modelling uncertainties}	\\						
\hline							
Fact. and renorm. scales		&$	0.004	$&$	0.005	$	\\
Top-quark mass		&$	0.001	$&$	0.001	$	\\
Parton distribution functions		&$	{<}0.001	$&$	{<}0.001	$	\\
$\tau$-lepton decay		&$	{<}0.001	$&$	{<}0.001	$	\\
Top-quark \pt\ reweighting	&$	0.001	$&$	{<}0.001	$	\\
\hline														
Unfolding		&$	0.006	$&$	0.001	$	\\							
\hline							
Total systematic uncertainty		&$	0.008	$&$	0.006	$	\\
\hline							
\end{tabular}
\end{table}

\section{Results}
\label{sec:results}

\begin{figure}[hbt]
\begin{center}
\includegraphics[width=0.49\linewidth]{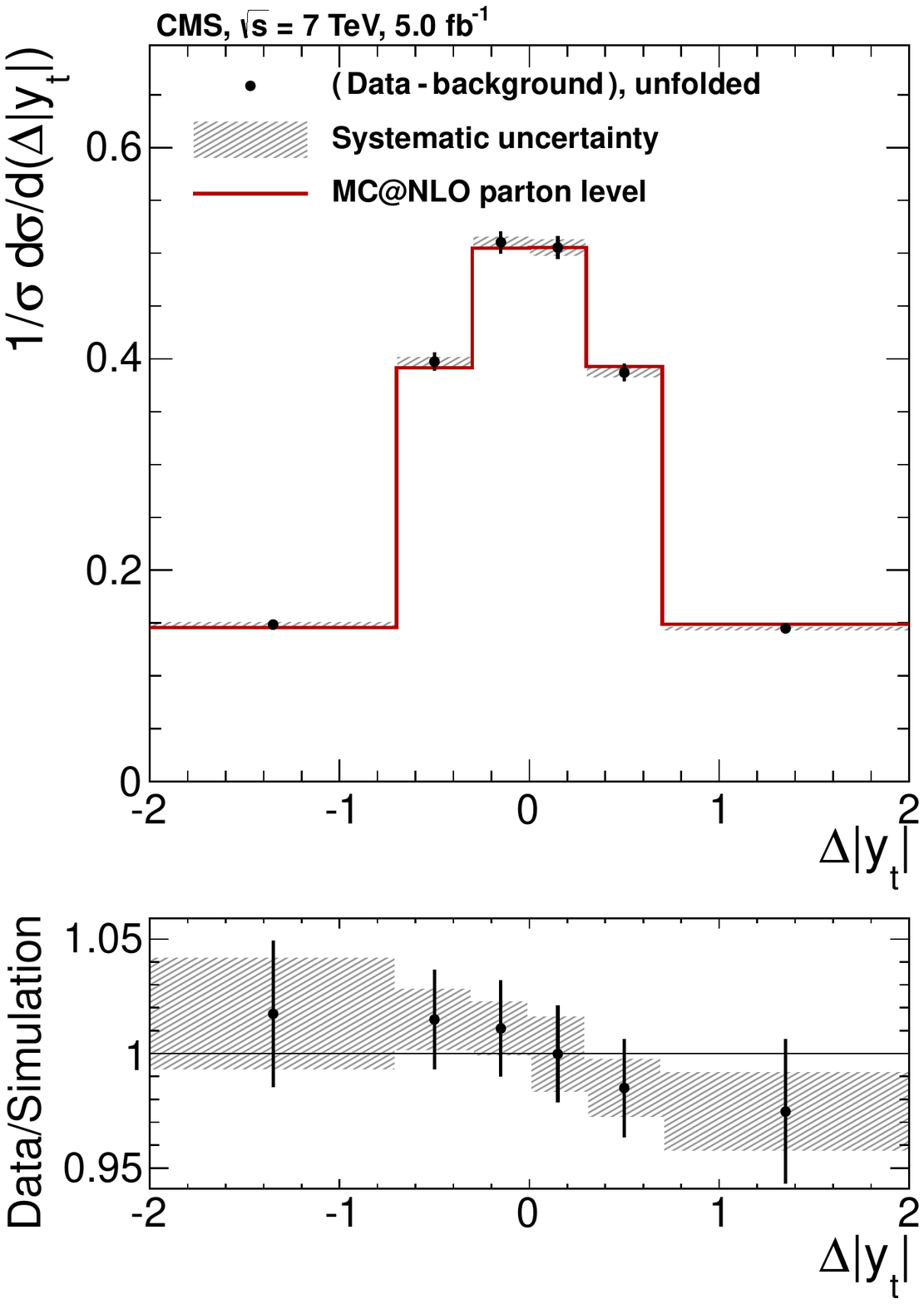}
\includegraphics[width=0.49\linewidth]{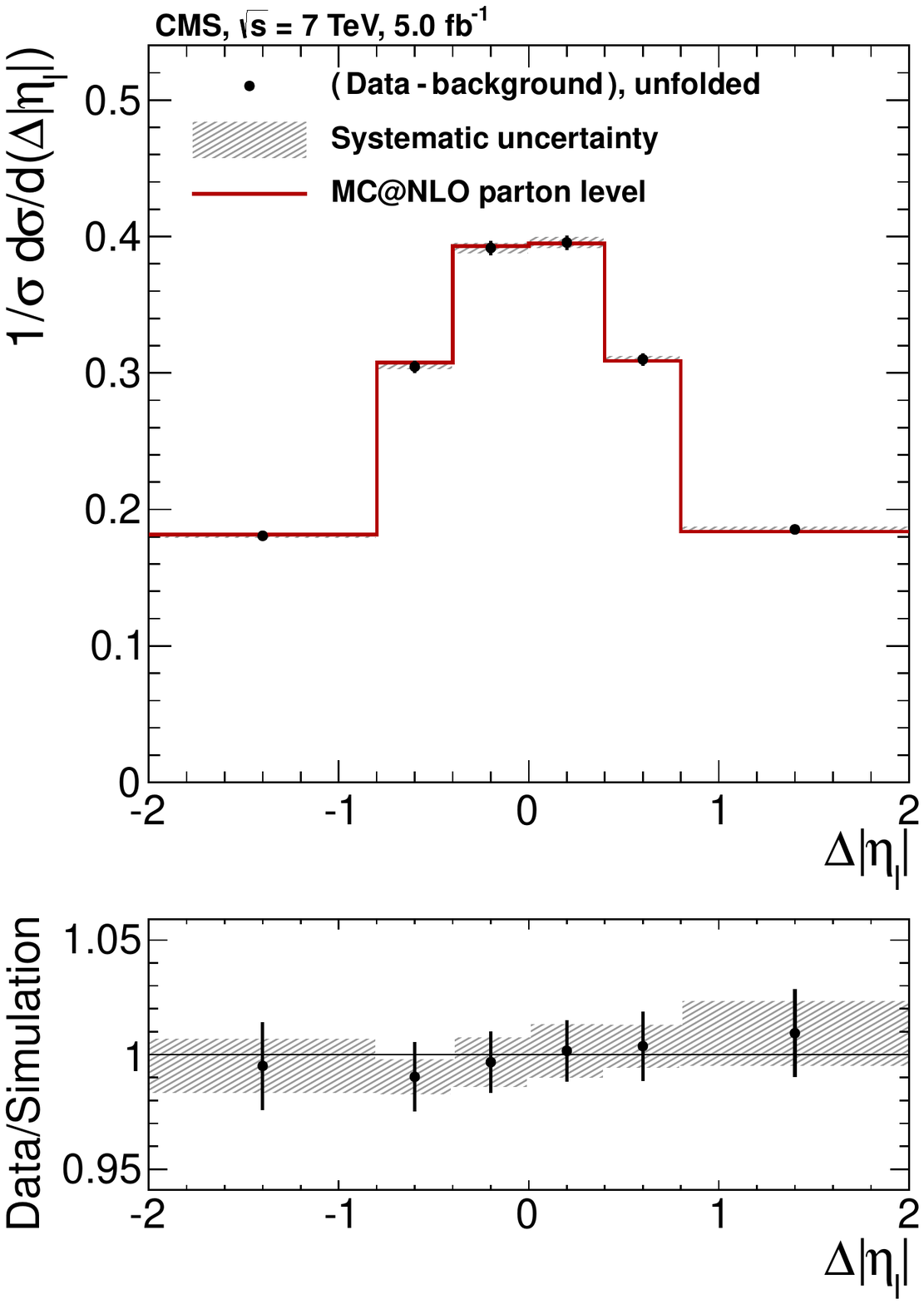}
\caption{\label{fig:ResultsUnfolded}
Top: Background-subtracted and unfolded differential measurements of $\Delta \abs{y_\cPqt}$ (left) and $\Delta \abs{\eta_\ell}$ (right), both normalised to unit area (points), and the parton-level predictions from \MCATNLO\ (histograms).
Bottom: The ratio between the data and the \MCATNLO\ prediction for $\Delta \abs{y_\cPqt}$ (left) and $\Delta \abs{\eta_\ell}$ (right).
The error bars represent the statistical uncertainties in the data, while the systematic uncertainties are represented by the hatched band.
The first and last bins include underflow and overflow events, respectively.
}
\end{center}
\end{figure}

The background-subtracted, unfolded, and normalised $\Delta \abs{y_\cPqt}$ and $\Delta \abs{\eta_\ell}$ distributions for the selected data events are shown in \fig~\ref{fig:ResultsUnfolded}, along with the parton-level predictions obtained with the \MCATNLO\ generator. The measured and predicted values are consistent.

The measured values of $A_\mathrm{C}$ and $A^\text{lep}_\mathrm{C}$, unfolded to the parton level, are presented in \tab~\ref{tab:ResultsUnfolded}, where they are
compared to the predictions from the \MCATNLO\ \ttbar\ sample and from NLO calculations~\cite{ref:BernChargeAsym}.
Correlations between the contents of different bins, introduced by the unfolding process, are accounted for in the calculation of the uncertainties.
The measured values are consistent with the expectations of the SM.

\begin{figure}[hbt]
\centering
\includegraphics[width=0.49\linewidth]{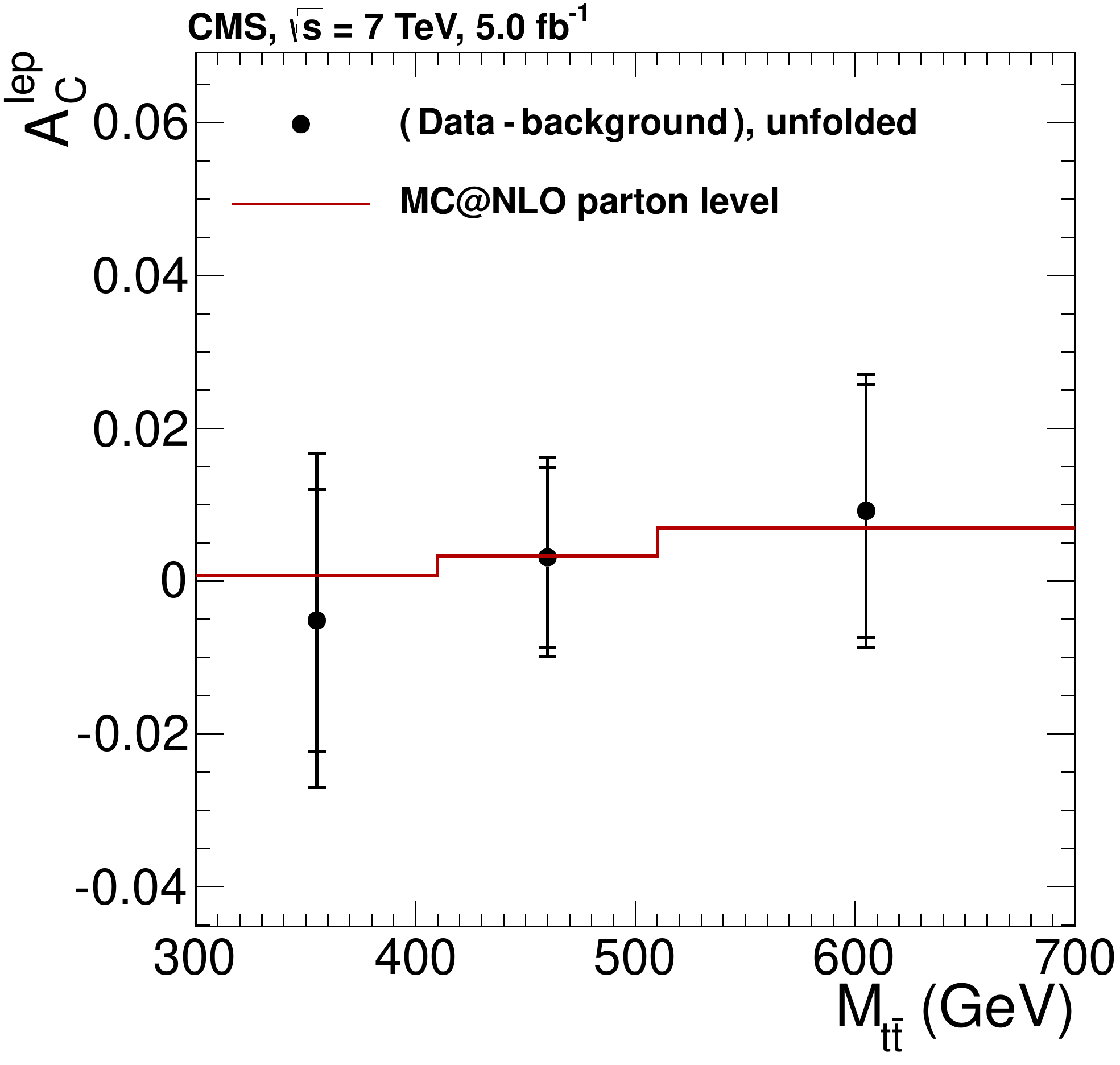}
\includegraphics[width=0.49\linewidth]{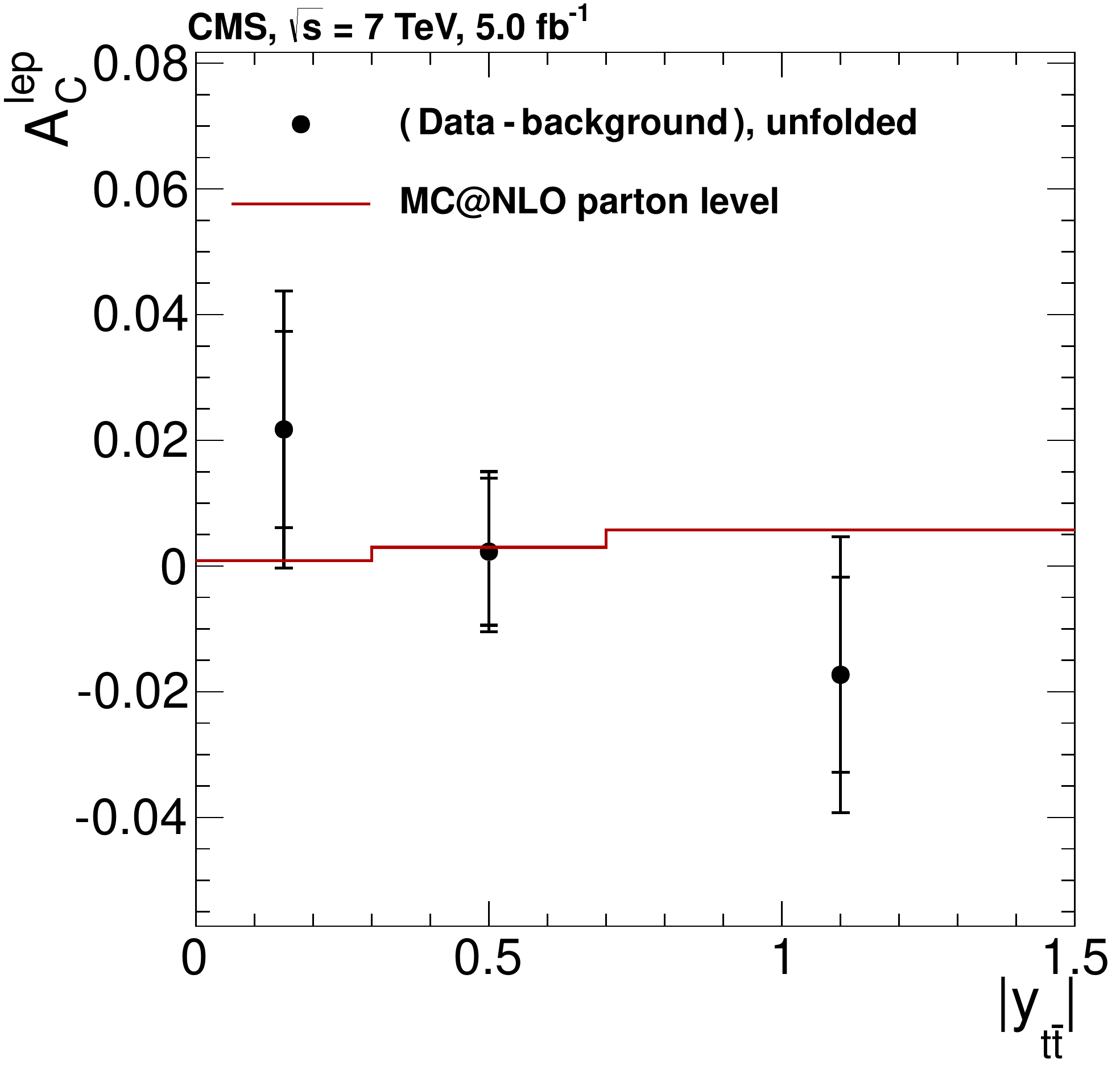}
\includegraphics[width=0.49\linewidth]{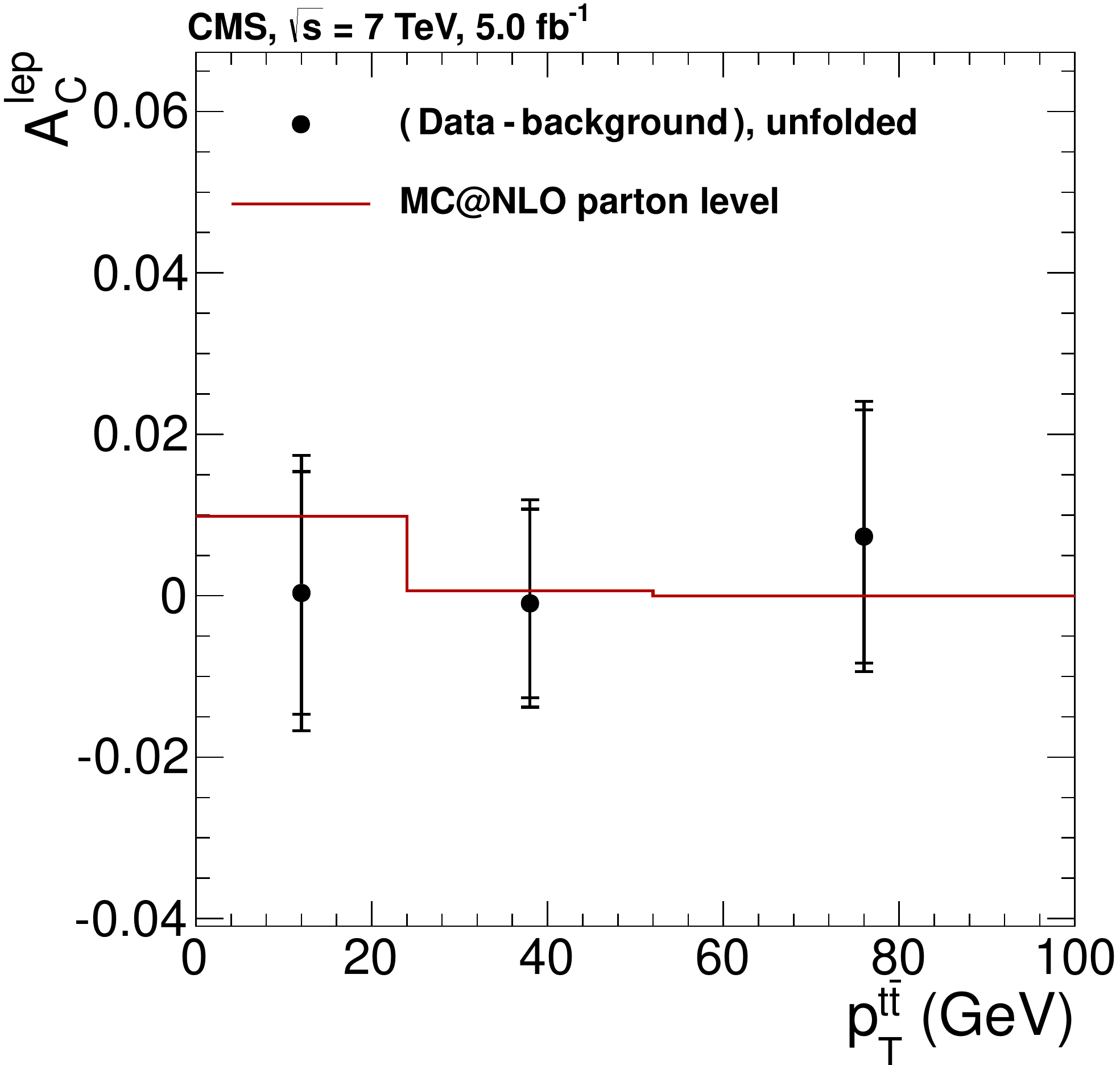}
\caption{\label{fig:AlepCdiff}\protect
Dependence of the unfolded $A^\text{lep}_\mathrm{C}$ values (points) on $M_{\ttbar}$ (top left), $\abs{y_{\ttbar}}$ (top right), and $\pt^{\ttbar}$ (bottom), and the parton-level predictions from \MCATNLO\ (histograms).
The inner and outer error bars represent the statistical and total uncertainties, respectively.
The last bin of each plot includes overflow events.
}
\end{figure}

We also measure the dependence of the unfolded $A^\text{lep}_\mathrm{C}$ values on
$M_{\ttbar}$, $\abs{y_{\ttbar}}$, and $\pt^{\ttbar}$.
To do so, we apply the same unfolding procedure
on a two-dimensional distribution consisting of two bins in $\Delta \abs{\eta_\ell}$ ($\Delta \abs{\eta_\ell}>0$ and $\Delta \abs{\eta_\ell}<0$) and three bins in
$M_{\ttbar}$, $\abs{y_{\ttbar}}$, or $\pt^{\ttbar}$. Since the regularisation procedure makes use of the second-derivative matrix, which is not well-defined for a two-bin distribution,
the regularisation constraint is applied only along the $M_{\ttbar}$, $\abs{y_{\ttbar}}$, and $\pt^{\ttbar}$ coordinates (this method was used previously in \reference~\cite{ref:CDFafb2013}).
The dependencies of the unfolded $A^\text{lep}_\mathrm{C}$ measurements on $M_{\ttbar}$, $\abs{y_{\ttbar}}$, and $\pt^{\ttbar}$ are shown in \fig~\ref{fig:AlepCdiff}.
The corresponding values of $A^\text{lep}_\mathrm{C}$ are given in \tab~\ref{tab:Results2DUnfolded}.
The results are consistent with the \MCATNLO\ predictions.
We did not measure the differential $A_\mathrm{C}$ values by this method, because the large migration of events between positive and negative $\Delta \abs{y_\cPqt}$ was found to result in a biased response when only two bins in $\Delta \abs{y_\cPqt}$ were used for the unfolding.

\begin{table}[!htpb]
\centering
\topcaption{\label{tab:ResultsUnfolded}
The unfolded $A_\mathrm{C}$ and $A^\text{lep}_\mathrm{C}$ measurements and parton-level predictions from the \MCATNLO\ simulation and from NLO calculations~\cite{ref:BernChargeAsym}.
For the data, the first uncertainty is statistical and the second is systematic. For the simulated results, the uncertainties are statistical only,
while the uncertainties in the NLO calculations come from varying the factorisation and renormalisation scales up and down by a factor of two.
}
\begin{tabular}{l|rcc}
\hline
Variable  &   \multicolumn{1}{c}{Data (unfolded)} &  \multicolumn{1}{c}{\MCATNLO\ prediction} & \multicolumn{1}{c}{NLO theory} \\
\hline
& & \\ [-2.2ex]
$A_\mathrm{C}$ &   $-0.010 \pm 0.017 \pm 0.008$  &  $ 0.004 \pm 0.001$  & $0.0123 \pm 0.0005$ \\ [0.3ex]
$A^\text{lep}_\mathrm{C}$ & $0.009 \pm 0.010 \pm 0.006$  &  $ 0.004 \pm 0.001$  & $0.0070 \pm 0.0003	$  \\ [0.3ex]
\hline
\end{tabular}
\end{table}

\begin{table}[!htpb]
\centering
\topcaption{\label{tab:Results2DUnfolded}
Measurements of the unfolded $A^\text{lep}_\mathrm{C}$ values in bins of $M_{\ttbar}$, $\abs{y_{\ttbar}}$, and $\pt^{\ttbar}$, and the parton-level predictions from \MCATNLO.
For the data, the first uncertainty is statistical and the second is systematic. For the predictions, the uncertainties are statistical only.
}
\begin{tabular}{r|rc}
\hline & & \\ [-2.2ex]
\multicolumn{1}{c|}{Bin} &   \multicolumn{1}{c}{Data (unfolded)} &  \multicolumn{1}{c}{\MCATNLO\ prediction} \\ [0.3ex]
\hline
& & \\ [-2.2ex]
 $M_{\ttbar} < 410$\GeV & $ -0.005 \pm 0.017 \pm 0.013$  &  $0.001 \pm 0.008$   \\
 $410$ $\le M_{\ttbar} < 510$\GeV & $ 0.003 \pm 0.012 \pm 0.006$  &  $0.003 \pm 0.007$   \\
 $M_{\ttbar} \ge 510$\GeV & $ 0.009 \pm 0.017 \pm 0.007$  &  $0.007 \pm 0.008$   \\ [0.3ex]
\hline
 & & \\ [-2.2ex]
 $\abs{y_{\ttbar}} < 0.3$~~~~~~~~~\, & $0.022 \pm 0.016 \pm 0.016$  &  $0.001 \pm 0.009$   \\
 $0.3 \le \abs{y_{\ttbar}} < 0.7$~~~~~~~~~\, & $0.002 \pm 0.012 \pm 0.005$  &  $0.003 \pm 0.008$   \\
 $\abs{y_{\ttbar}} \ge 0.7$~~~~~~~~~\, & $-0.017 \pm 0.016 \pm 0.015$  &  $0.006 \pm 0.006$   \\ [0.3ex]
\hline
 & & \\ [-2.2ex]
 $\pt^{\ttbar} < 24$\GeV~~\: & $0.000 \pm 0.015 \pm 0.008$  &  $0.009 \pm 0.007$   \\
 $24$ $\le \pt^{\ttbar} < 52$\GeV~~\: & $-0.001 \pm 0.012 \pm 0.005$  &  $0.000 \pm 0.008$   \\
 $\pt^{\ttbar} \ge 52$\GeV~~\: & $0.007 \pm 0.016 \pm 0.006$  &  $0.000 \pm 0.007$   \\ [0.3ex]
\hline
\end{tabular}
\end{table}

\section{Summary}

The first measurements in the dilepton final state of the difference in the $\abs{y}$ distributions of top quarks and antiquarks
and in the $\abs{\eta}$ distributions of positive and negative leptons have been presented,
in terms of the asymmetry variables $A_\mathrm{C}$ and $A^\text{lep}_\mathrm{C}$, respectively.
The data sample of \ttbar\ events
corresponds to a total integrated luminosity of 5.0\fbinv from pp collisions at $\sqrt{s}=7$\TeV, collected by the CMS experiment at the LHC.
The measured value of $A_\mathrm{C}$ is  $-0.010 \pm 0.017\stat \pm 0.008\syst$ and of $A^\text{lep}_\mathrm{C}$ is $0.009 \pm 0.010\stat \pm 0.006\syst$, both unfolded to the parton level.
The differential distributions of $A^\text{lep}_\mathrm{C}$ as a function of the \ttbar\ system variables $M_{\ttbar}$, $\abs{y_{\ttbar}}$, and $\pt^{\ttbar}$ have also been determined.
All measurements are found to be in agreement with standard model expectations,
and can help constrain theories beyond the standard model~\cite{ref:Krohn}.

\section*{Acknowledgements}
We congratulate our colleagues in the CERN accelerator departments for the excellent performance of the LHC and thank the technical and administrative staffs at CERN and at other CMS institutes for their contributions to the success of the CMS effort. In addition, we gratefully acknowledge the computing centres and personnel of the Worldwide LHC Computing Grid for delivering so effectively the computing infrastructure essential to our analyses. Finally, we acknowledge the enduring support for the construction and operation of the LHC and the CMS detector provided by the following funding agencies: BMWF and FWF (Austria); FNRS and FWO (Belgium); CNPq, CAPES, FAPERJ, and FAPESP (Brazil); MES (Bulgaria); CERN; CAS, MoST, and NSFC (China); COLCIENCIAS (Colombia); MSES and CSF (Croatia); RPF (Cyprus); MoER, SF0690030s09 and ERDF (Estonia); Academy of Finland, MEC, and HIP (Finland); CEA and CNRS/IN2P3 (France); BMBF, DFG, and HGF (Germany); GSRT (Greece); OTKA and NIH (Hungary); DAE and DST (India); IPM (Iran); SFI (Ireland); INFN (Italy); NRF and WCU (Republic of Korea); LAS (Lithuania); MOE and UM (Malaysia); CINVESTAV, CONACYT, SEP, and UASLP-FAI (Mexico); MBIE (New Zealand); PAEC (Pakistan); MSHE and NSC (Poland); FCT (Portugal); JINR (Dubna); MON, RosAtom, RAS and RFBR (Russia); MESTD (Serbia); SEIDI and CPAN (Spain); Swiss Funding Agencies (Switzerland); NSC (Taipei); ThEPCenter, IPST, STAR and NSTDA (Thailand); TUBITAK and TAEK (Turkey); NASU (Ukraine); STFC (United Kingdom); DOE and NSF (USA).

Individuals have received support from the Marie-Curie programme and the European Research Council and EPLANET (European Union); the Leventis Foundation; the A. P. Sloan Foundation; the Alexander von Humboldt Foundation; the Belgian Federal Science Policy Office; the Fonds pour la Formation \`a la Recherche dans l'Industrie et dans l'Agriculture (FRIA-Belgium); the Agentschap voor Innovatie door Wetenschap en Technologie (IWT-Belgium); the Ministry of Education, Youth and Sports (MEYS) of Czech Republic; the Council of Science and Industrial Research, India; the Compagnia di San Paolo (Torino); the HOMING PLUS programme of Foundation for Polish Science, cofinanced by EU, Regional Development Fund; and the Thalis and Aristeia programmes cofinanced by EU-ESF and the Greek NSRF.

\bibliography{auto_generated}   

\cleardoublepage \appendix\section{The CMS Collaboration \label{app:collab}}\begin{sloppypar}\hyphenpenalty=5000\widowpenalty=500\clubpenalty=5000\textbf{Yerevan Physics Institute,  Yerevan,  Armenia}\\*[0pt]
S.~Chatrchyan, V.~Khachatryan, A.M.~Sirunyan, A.~Tumasyan
\vskip\cmsinstskip
\textbf{Institut f\"{u}r Hochenergiephysik der OeAW,  Wien,  Austria}\\*[0pt]
W.~Adam, T.~Bergauer, M.~Dragicevic, J.~Er\"{o}, C.~Fabjan\cmsAuthorMark{1}, M.~Friedl, R.~Fr\"{u}hwirth\cmsAuthorMark{1}, V.M.~Ghete, C.~Hartl, N.~H\"{o}rmann, J.~Hrubec, M.~Jeitler\cmsAuthorMark{1}, W.~Kiesenhofer, V.~Kn\"{u}nz, M.~Krammer\cmsAuthorMark{1}, I.~Kr\"{a}tschmer, D.~Liko, I.~Mikulec, D.~Rabady\cmsAuthorMark{2}, B.~Rahbaran, H.~Rohringer, R.~Sch\"{o}fbeck, J.~Strauss, A.~Taurok, W.~Treberer-Treberspurg, W.~Waltenberger, C.-E.~Wulz\cmsAuthorMark{1}
% ...
\vskip\cmsinstskip
% ...
1:~~Also at Vienna University of Technology, Vienna, Austria\\
2:~~Also at CERN, European Organization for Nuclear Research, Geneva, Switzerland\\
% ...

\end{sloppypar}
\end{document}